\documentclass[aps,preprint]{revtex4}%
\usepackage{amsfonts}
\usepackage{amsmath}
\usepackage{amssymb}
\usepackage{graphicx}%
\begin{document}
\preprint{ }
\title[Potential curves for Ar$_{2}$]{Potential energy curves and transition moments for the excimer states
0$_{u}^{+}$ ($^{3}$P$_{1}$) and 1$_{u}$ ($^{3}$P$_{2}$) of Ar$_{2}$}
\author{Matti Selg}
\affiliation{Institute of Physics of the University of Tartu, Riia 142, 51014 Tartu, Estonia}
\pacs{PACS number}

\begin{abstract}
Exactly solvable rererence potentials of several smoothly joined Morse-type
components were constructed for the lowest two excimer states of Ar$_{2}$
molecule. The parameters of the potentials have been ascertained by fitting to
the experimental data, and they are reliable in a wide range of nuclear
separations ($r\geq1.9$ \AA ). A large number of quantum mechanical
Franck-Condon factors for the 0$_{u}^{+}\rightarrow$ 0$_{g}^{+}$ and
1$_{u}\rightarrow$ 0$_{g}^{+}$ transitions have been calculated and compared
with the observed spectroscopic features. The fitting procedure also involved
dipole transition moments, which have been adjusted to the known radiative
lifetimes of the vibrational levels. The resulting potential energy curves
accurately reproduce the first and the second emission continua of
Ar$_{2}^{\ast}$ as well as the oscillatory spectrum related to their inner
turning point region. The numbering and the positions of the vibrational
levels reported by Herman \textit{et al. }[P. R. Herman, P. E. LaRocque, and
B. P. Stoicheff, J. Chem. Phys. \textbf{89}, 4535 (1988)] have been confirmed.
The calculated spectroscopic parameters, $\omega_{e}=296.26$ cm$^{-1},$
$D_{e}=6128.3$ cm$^{-1}$ for the 0$_{u}^{+}$ state, and $\omega_{e}=287.30$
cm$^{-1},$ $D_{e}=5929.6$ cm$^{-1}$ for the 1$_{u}$ state, are consistent with
previous estimations. Separation between the minima of the potential wells at
$R_{e}=2.3893$ \AA \ was found to be 50.60 meV, compared with 75.24 meV
splitting in the separated-atom limit.

\end{abstract}
\maketitle

\section{Introduction}

Rare gases (RG) are important laser media in the vacuum ultraviolet (VUV)
spectral region \cite{Hutchinson, McCusker}, and this is one of the reasons
for the undying interest in these chemically inactive species. Normally, they
are almost atomic gases, because electronically unexcited RG dimers are only
weakly bound (except radioactive Rn$_{2}$ with dissociation energy of about
400 K \cite{Runeberg}). In contrast to this, the electronically excited RG
dimers (excimers) in their lowest states 1$_{u}$ and 0$_{u}^{+}$ are strongly
bound. These excimers can be created with the help of various excitation
sources (charged particle beams, synchrotron radiation, etc.), and their decay
produces photons in the VUV region, forming two intense continua observed long
ago for all RGs (except Rn) \cite{Ar-1972, He-1973, Ne-1973, Kr-1974, Xe-1976}
(see also \cite{Becker-2002, Ulrich-2004} for the recent Ne and He data).

A vast literature exists on the RG dimers in their ground electronic state
\cite{Runeberg, Aziz-90, Aziz-93, Woon-94, Aziz-97, Cybulski, Faas, Tang,
Slavicek}, because they are good prototypes for studying van der Waals
interactions. At the same time, there is still a lack of knowledge about the
RG excimer states, which are of crucial importance for the development of
efficient light sources in the VUV region \cite{Ulrich-2004, Wieser-97,
APL-98, PP-2000, APL-2001, PRA-2003}. Of special interest in this context are
the heavier excimers for which the laser emission has been reported
\cite{Hutchinson, McCusker}. The observed laser effect is related to the
so-called second continua, broad structureless emission bands with maxima at
126, 146 and 172 nm for Ar$_{2}^{\ast}$, Kr$_{2}^{\ast}$, and Xe$_{2}^{\ast}$,
respectively. They correspond to transitions from the vibrationally relaxed
1$_{u}$ and 0$_{u}^{+}$ states to the repulsive part of the ground electronic
state 0$_{g}^{+}$, in contrast to the so-called first continua originating
from the highest vibrational levels of the 0$_{u}^{+}$ excimer state, and
therefore being spectrally close to the atomic $^{3}$P$_{1}\rightarrow$ $^{1}%
$S$_{0}$ resonance lines. The first continuum is related to the excimers'
classical outer turning point region, while the weak oscillatory emission from
the region of their inner turning points has also been observed under
selective synchrotron radiation excitation \cite{Moeller}.

All these important spectroscopic features can be uniformly treated
\cite{Selg-2003, Selg-2007}, but to this end, one needs reliable potential
energy curves (PECs) and coordinate dependencies of the related transition
moments, as well as a reasonable model to describe vibrational relaxation of
the excimers. In this paper, we concentrate on the first two issues, while the
relaxation dynamics and the time-dependent emission spectra will be studied in
a separate paper \cite{Selg-Future}. Valuable information on the energetic
position, absolute numbering and radiative lifetimes of the higher vibrational
levels for all heavier RG excimers has been obtained in a series of supersonic
jet expansion experiments \cite{Stoicheff-Xe, Stoicheff-Kr, Stoicheff-Ar,
Madej}, using tunable VUV radiation to produce high-resolution fluorescence
excitation spectra. Unfortunately, these high-quality reference data concern
only a narrow range of vibrational levels near the dissociation limit, and one
cannot construct reliable PECs for RG excimers solely on this basis. Various
\textit{ab initio} and semi-empirical potentials are available
\cite{Castex-Xe, Castex-Ar, Gadea, Yates, Messing, Audouard, Jonin}, which
satisfactorily reproduce some of the characteristic spectroscopic features,
but not the whole variety of available experimental data. Apart from
imperfections of the PECs of the excimer states, the discrepancies between
theoretical and experimental spectra may be caused by inaccuracies of the
ground state potentials in the short distance range.

Thus, the spectroscopic properties of RG dimers certainly need further
investigation. This is the motivation for the present study, the main goal of
which is to construct accurate reference potentials of relatively simple
analytic form, by directly fitting their parameters to the observed data, and
taking account of the relevant theoretical considerations. In principle, the
approach that will be described below can be applied to any RG dimers, since
they have much in common. For several reasons, however, only Ar$_{2}$
molecules are under examination in this paper. First, compared with Xe and Kr,
the effects of spin-orbit coupling are much less pronounced in Ar, which makes
the theoretical analysis more simple and, hopefully, more adequate. Second, a
sufficient amount of good-quality experimental data are available for Ar gas,
which, again, is an advantage compared with Xe and Kr. Third, there is a
practical point of special interest in Ar$_{2}^{\ast}$ excimers, as they produce
the highest energy photons (9.8 eV) for the potential laser applications, which
is the field of great interest and permanent development \cite{Neeser, Tanaka}.
It therefore seems reasonable to test the method on Ar$_{2}$ molecules, leaving
more complex cases for the future analysis.

The reference potential approach to different spectral problems has been
described in detail elsewhere (see, e.g., \cite{Selg-2003, Selg-2006} and
references therein). In this paper, we are going to construct smooth
multi-component exactly solvable reference potentials for the lowest excimer
states 1$_{u}$ and 0$_{u}^{+}$, and for the ground electronic state 0$_{g}%
^{+}$ of the Ar$_{2}$ molecule. These approximate PECs should be realistic in
a wide range of nuclear separations related to the observed spectroscopic
features. To achieve this goal, fully quantum mechanical calculations of
bound-bound and bound-free Franck-Condon factors for 1$_{u}$ $\rightarrow$
0$_{g}^{+}$ and 0$_{u}^{+}$ $\rightarrow$ 0$_{g}^{+}$ transitions have been
performed. The procedure has to be repeated many times, varying the parameters
of the PECs and the transition moments. In this way, step by step, the
calculated energy levels and their radiative lifetimes, as well as the
intensity patterns of Franck-Condon factors, can be fitted to the available
experimental data.

The paper is organized as follows. In Section II, a short overview of the
theoretical background of the method is given, specific properties of the
reference potentials for the 0$_{g}^{+}$, 0$_{u}^{+}$ and 1$_{u}$ electronic
states of Ar$_{2}$ are described (in sub-sections A, B, and C, respectively),
and the details of the fitting procedure are explained. Section III is devoted
to demonstrating the calculated Franck-Condon factors, transition
probabilities and radiative lifetimes for the vibrational levels of the
0$_{u}^{+}$ and 1$_{u}$ states. Finally, a brief discussion of the obtained
results and future prospects of the research are given in Section IV, which
concludes the paper.

\section{Exactly solvable multi-component reference potentials}

At first sight, the reference potential approach we are going to adopt may
seem naive and unjustified. However, the solution to a problem always
depends on the context. Modern spectroscopic analysis often involves many
thousand items of high resolution input data, but unfortunately, such
detailed information is not yet available neither for the ground state of Ar$%
_{2}$ molecule nor for the excimer states 1$_{u}$ and 0$_{u}^{+}$. Indeed,
the scattering data for these species are almost lacking. Moreover, even the
total number of the bound states is not conclusively established, and only a
narrow range of discrete levels has been accurately ascertained for the excimer
states \cite{Stoicheff-Ar}, while their absolute numbering still needs to be
confirmed. Under these circumstances, there is no chance to deduce the "real"
PECs directly from the experimental data. One can only take some steps towards
this goal by combining the available experimental data with the relevant
theoretical results. In this context, the simple but mathematically rigorous
approach in question has proven to be quite useful.

To begin with the analysis, let us recall another important peculiarity of
the system under study. Namely, for all RGs, there is a big difference
(about 1.4 \AA\ in the case of Ar$_{2}$)\ between the minima of the tiny
potential well of the ground state and those for the excimer states. It
means that bound-free transitions to the repulsive wall of the ground state
PEC are responsible for the major part of the observed fluorescence spectra.
Of course, as we demonstrate, bound-bound transitions are also very
important to reveal the details of the first emission continuum.
Unfortunately, this fine structure, which should contain tens of spectral
lines in the range of less than 0.1 eV width (compared with more than 4.5 eV
width of the overall spectrum), has not yet been detected experimentally.
Moreover, the spectral resolution in fluorescence experiments on Ar and other
RGs is typically $\sim $10 meV. Therefore, at the present level of our
knowledge, one should not put too much effort into deducing the PECs that would
reproduce the observed (relatively few) level positions with the utmost
accuracy. Formally, one can always achieve this specific goal, but this does not
guarantee the reliability of the potentials in the major part of the domain.
Indeed, as is well known, even the full knowledge of the discrete energy
spectrum combined with the full scattering information is insufficient to
uniquely determine the potential \cite{Chadan}.

In view of the above, we set the following criteria to the quality of the
reference potentials for the Ar$_{2}^{\ast }$ excimer states: 1) these PECs
should reproduce the observed bound-bound transition energies \cite%
{Stoicheff-Ar} with $\sim $1 meV accuracy; 2) they should reproduce all
details of the fluorescence spectra in the range from about 7 to 11.63 eV
with $\sim $10 meV accuracy. To achieve these goals, we use a fitting method
which is based on repeated and accurate calculation of the Franck-Condon
factors. Naturally, this strictly quantum mechanical procedure presumes
highly efficient solution of the Schr\"{o}dinger equation, because this
elementary act has to be performed many million times within a reasonable
time scale. For this reason, we try to adopt an analytic approach of calculating
the energy eigenfunctions, which indirectly means that we have to construct
exactly solvable approximants (reference potentials) to the "real" PECs. If one
sets a "physical constraint", requiring continuity of the potential and its
first derivative in the whole physical domain, the number of suitable options
becomes quite limited. Certainly one of the best choices is to compose a
reference potential of several Morse-type \cite{Morse} pieces%
\begin{equation}
V(r)=V_{k}+D_{k}\left[  \exp(-\alpha_{k}(r-R_{k}))-1\right]  ^{2},\qquad
r\in(0,\infty),
\end{equation}
where $V_{k},$ $D_{k},$ $\alpha_{k}$ and $R_{k}$ are real (not definitely
positive!) constants, and the subscript $k=0,1,2...$labels different
components smoothly joined at the boundary points $X_{k+1}$.

As is well known, the classical Morse potential belongs to the family of shape
invariant potentials \cite{SUSY}, and its energy eigenvalue problem can be
solved with the help of solely algebraic techniques. The shape invariance is
lost, if there are several analytically different components, as assumed in
Eq. (1). Consequently, the discrete energy levels of such multi-component
potentials cannot be given in an explicit analytic form. With some concession,
we can still preserve the term "exactly solvable" for this kind of piece-wise
potentials, because the two linearly independent solutions of the related
Schr\"{o}dinger equation can be always found analytically to any desired
accuracy. Indeed, introducing a dimensionless variable $y_{k}\equiv2a_{k}%
\exp(-\alpha_{k}(r-R_{k})),$ the Schr\"{o}dinger equation for a Morse-type PEC
can be converted into a confluent hypergeometric form \cite{Bateman}%
\begin{equation}
y_{k}\frac{d^{2}G(a_{k},\mu_{k};y_{k})}{dy_{k}^{2}}+(2\mu_{k}+1-y_{k}%
)\frac{dG(a_{k},\mu_{k};y_{k})}{dy_{k}}+(a_{k}-\mu_{k}-1/2)G(a_{k},\mu
_{k};y_{k})=0.
\end{equation}
Here $a_{k}\equiv\sqrt{D_{k}/C}/\alpha_{k},C\equiv\frac{\hbar^{2}}{2m}$ is a
characteristic constant ($C=0.10464$ meV$\cdot$\AA $^{2}$ for Ar$_{2}$),
$\mu_{k}\equiv\frac{\sqrt{(V_{k}+D_{k}-E)/C}}{\alpha_{k}},$ and the solutions
of the Schr\"{o}dinger equation read (up to normalization) $\Psi=\exp
(-y_{k}/2)y_{k}^{\mu_{k}}G(a_{k},\mu_{k};y_{k}).$ Note that the parameters
$a_{k},\mu_{k}$ and the coordinate $y_{k}$ may be imaginary, if $D_{k}<0$ or
the total energy $E>V_{k}+D_{k}.$

There are several possibilities to construct the fundamental system of
solutions for Eq. (2) (see \cite{Bateman}, Chapter 6, for a thorough
overview). For example, one can make use of the special solutions%
\begin{equation}
G_{1}=\Phi(-a_{k}+\mu_{k}+1/2,2\mu_{k}+1;y_{k}),\text{ }G_{2}=y_{k}^{-2\mu
_{k}}\Phi(-a_{k}-\mu_{k}+1/2,-2\mu_{k}+1;y_{k}),
\end{equation}
where the symbols%
\begin{equation}
\Phi(a,c;x)\equiv1+\frac{a\cdot x}{1!\cdot c}+\frac{a(a+1)\cdot x^{2}}{2!\cdot
c(c+1)}+...
\end{equation}
denote confluent hypergeometric functions. Correspondingly, the two linearly
independent solutions of the original Schr\"{o}dinger equation read%
\begin{equation}
\Psi_{k1}=y_{k}^{\mu_{k}}S(a_{k},\mu_{k};y_{k}),\text{ }\Psi_{k2}=y_{k}%
^{-\mu_{k}}S(a_{k},-\mu_{k};y_{k}),
\end{equation}
where we have introduced another very useful function%
\begin{equation}
S(a,c;x)\equiv\exp(-x/2)\Phi(-a+c+1/2,2c+1;x),
\end{equation}
which can be evaluated as follows \cite{Tricomi}:%
\begin{gather}
S(a,c;x)=\sum_{n=0}^{\infty}B_{n},\text{ }B_{0}=1,\text{ }B_{1}=-\frac
{ax}{2c+1},\\
B_{n}=\frac{x}{n(2c+n)}(-aB_{n-1}+\frac{x}{4}B_{n-2}),\text{ }%
n=2,3,...\nonumber
\end{gather}
From Eqs. (5) to (7) one can infer that $\Psi_{k1}$ and $\Psi_{k2}$\ are
always complex conjugates, if $E>V_{k}+D_{k}.$

\subsection{Four-component reference potential for the ground state of
Ar$_{2}$}

After the brief theoretical introduction, let us try to put the ideas into
practice. Our fitting procedure does not actually involve the parameters of
the ground state, but only those of the excimer states 1$_{u}$ and 0$_{u}%
^{+}.$ In other words, the PEC for the ground state is assumed to be fixed.
Nevertheless, in order to solve the whole quantum mechanical problem exactly, we
have to construct an exactly solvable (in the above-mentioned sense) reference
potential for the ground state as well. Of course, such a constraint is
technical rather than conceptual. The Schr\"{o}dinger equation can be always
solved numerically, but in the present context the main problem is how to fix
a sufficiently realistic PEC. The shape of the ground state potential for
Ar$_{2}$\ is accurately known in the intermediate and long-distance range, but
the knowledge about its repulsive short-distance part is rather ambiguous. As
mentioned, this region is very important to interpret the observed
spectroscopic features, thus our aim is to describe the repulsive wall as
adequately as possible.

The resulting reference PEC for Ar$_{2}$\ is shown in Fig. 1, and its
parameters are given in Table 1. All four components have the simple analytic
form of Eq. (1), and their parameters have been ascertained from the least
squares fit to the \textit{ab initio} CCSD(T) daug-cc-pV5Z-33211 potential by
Fern\'{a}ndez and Koch \cite{Fernandez}. This particular PEC has been
preferred, because this is the only work to date where both the ground state
and the excimer state (0$_{u}^{+}$) potentials have been calculated on the
same theoretical basis. These PECs nicely reproduce the observed spectra
related to the excimers' inner turning point region \cite{Moeller}, which is a
strong argument for their reliability at short distances. As can be seen in
Figs. 1 and 2, the ground state PEC by Fern\'{a}ndez and Koch is in good
agreement with the other available \textit{ab initio} potentials
\cite{Slavicek, Barker}. The reference potential for Ar$_{2}$ is almost
indistinguishable from the original PEC on the scale used in these figures.
Therefore, it is expected to be reliable enough for our purposes.

Now, let us have a look at the structure of this reference PEC. The two most
internal components ($r\leq X_{1}$ and $r\in\lbrack X_{1},X_{2}]$,
respectively) represent the so-called pseudo-Morse (PM) potentials
\cite{Selg-2003, Selg-2006} smoothly joined at the boundary point $X_{1}$.
Their important peculiarity is that the parameters $D_{k}$ and $\alpha_{k}$
($k=0,1$) are not independent, but $D_{k}=\frac{1}{4}C\alpha_{k}^{2}$. It
means that the tiny potential well is just of the limit depth to entirely lose
the discrete energy spectrum. Since $a_{k}=1/2$, and the pseudo-Morse
approximation is used only in the region where $E>V_{k}+D_{k},$ the two
special solutions $\Psi_{k1}$ and $\Psi_{k2}$ of the Schr\"{o}dinger equation
are complex conjugates. According to Eqs. (5) to (7),%
\begin{equation}
\Psi_{k1}=y_{k}^{i\beta_{k}}S(1/2,i\beta_{k};y_{k})=A_{k}(y_{k})e^{iB_{k}%
(y_{k})}e^{-i\alpha_{k}\beta_{k}(r-r_{k})},
\end{equation}
where
\begin{gather}
A_{k}(y_{k})e^{iB_{k}(y_{k})}\equiv1-\frac{y_{k}/4}{i\beta_{k}+1/2}\\
+\frac{\left(  y_{k}/4\right)  ^{2}}{\left(  i\beta_{k}+1/2\right)  1!}\left(
1-\frac{y_{k}/4}{i\beta_{k}+3/2}\right)  +\frac{\left(  y_{k}/4\right)  ^{4}%
}{\left(  i\beta_{k}+1/2\right)  \left(  i\beta_{k}+3/2\right)  2!}\left(
1-\frac{y_{k}/4}{i\beta_{k}+5/2}\right)  +....,\nonumber
\end{gather}
and, consequently, the general solution reads%
\begin{equation}
\Psi_{k}(r)=N_{k}A_{k}(y_{k})\cos\left[  B_{k}(y_{k})+\varphi_{k}-\alpha
_{k}\beta_{k}r\right]  ,\text{ }k=0,1.
\end{equation}
Here $\beta_{k}\equiv\left\vert \mu_{k}\right\vert ,$ while the normalization
factor $N_{k}$ and the phase constant $\varphi_{k}$ should be determined from
the continuity requirements of the wave function and its derivative.

The phase constant $\varphi_{0}$ for the most internal PM component can be
always (in most cases quite easily) ascertained analytically \cite{Selg-2006},
taking account of the physical boundary condition $\Psi\rightarrow0$ as
$r\rightarrow0$. The next phase constant $\varphi_{1}$ can then be determined
from the boundary condition $\frac{\Psi_{0}^{\prime}(X_{1})}{\Psi_{0}(X_{1}%
)}=\frac{\Psi_{1}^{\prime}(X_{1})}{\Psi_{1}(X_{1})}.$ In all cases analyzed in
this paper the wave function's logarithmic derivative can be expressed in the
form%
\begin{equation}
\frac{\Psi_{k}^{\prime}(r)}{\Psi_{k}(r)}\equiv\alpha_{k}\left[  R_{k}%
(r)+\frac{S_{k}(r)\tan\varphi_{k}+T_{k}(r)}{U_{k}(r)\tan\varphi_{k}+W_{k}%
(r)}\right]
\end{equation}
where $R_{k},S_{k},T_{k},U_{k}$ and $W_{k}$ are some characteristic functions
that do not depend on phase constants. Thus, if $\varphi_{0}$ is known,
$\varphi_{1}$ can be easily determined.

The region $r\in\lbrack X_{2},X_{3}]$ (including the minimum of the PEC) is
approximated by an ordinary Morse (OM) potential ($k=2$). Again, since the
energy range $E<V_{2}+D_{2}$ is out of interest, the special solutions
$\Psi_{21}$ and $\Psi_{22}$ are complex conjugates, and the general solution
of the Schr\"{o}dinger equation becomes%
\begin{equation}
\Psi_{2}(r)=N_{2}\left[  C_{2}(r)\tan\varphi_{2}+D_{2}(r)\right]  ,
\end{equation}
where%
\begin{align*}
C_{2}(r)  &  \equiv\operatorname{Re}\left[  S(a_{2},i\beta_{2};y_{2}\right]
\sin(k_{2}r)-\operatorname{Im}\left[  S(a_{2},i\beta_{2};y_{2}\right]
\cos(k_{2}r),\\
D_{2}(r)  &  \equiv\operatorname{Re}\left[  S(a_{2},i\beta_{2};y_{2}\right]
\cos(k_{2}r)+\operatorname{Im}\left[  S(a_{2},i\beta_{2};y_{2}\right]
\sin(k_{2}r),
\end{align*}
$k_{2}\equiv\alpha_{2}\beta_{2},$ and $\beta_{2}\equiv\left\vert \mu
_{2}\right\vert .$ To ascertain the phase constant $\varphi_{2}$, one uses Eq.
(11) and applies the boundary condition $\frac{\Psi_{1}^{\prime}(X_{2})}%
{\Psi_{1}(X_{2})}=\frac{\Psi_{2}^{\prime}(X_{2})}{\Psi_{2}(X_{2})}$.

The most external region $r\geq X_{3}$ ($k=3$) is approximated by a reversed
Morse (RM) potential with a negative "dissociation energy" $D_{3}$. This might
seem unphysical, because there is actually no hump on the original PEC.
However, since the hump of the reference PEC is very small and located in the
long-distance range (see Table 1) where $V(r)\approx0$, this artificial effect
is nearly negligible for spectroscopic applications, while the analytic
treatment remains as simple as in previously studied cases. Of course, one can
introduce more such components and gradually shift the maximum to an arbitrarily
long distance (in principle, to infinity), thus practically eliminating the artificial barrier. In this paper, this physically motivated but tedious
procedure has not been undertaken, because the desired quality of the PECs (see
the criteria set in Section II) can already be achieved with the help of only
3-4 components.

A thorough analysis of the bound-states region $E<V_{3}+D_{3}=0$ has been given
elsewhere \cite{Selg-2001}. For the scattering states ($E>0$), the wave function
reads%
\begin{equation}
\Psi_{3}(r)=\frac{2C_{3}(r)}{\sqrt{\tan^{2}\varphi_{3}+1}}\{\cos\left[
D_{3}(r)-kr\right]  -\tan\varphi_{3}\cdot\sin\left[  D_{3}(r)-kr\right]  \},
\end{equation}
where $C_{3}(r)\exp\left[  iD_{3}(r)\right]  \equiv S(ia_{3},i\beta_{3}%
;ix_{3}),\beta_{3}\equiv\left\vert \mu_{3}\right\vert ,x_{3}\equiv\left\vert
y_{3}\right\vert ,$ and $k\equiv\alpha_{3}\beta_{3}=\sqrt{E/C}.$ Since
$C_{3}(r)\rightarrow1$ and $D_{3}(r)\rightarrow0$ as $r\rightarrow\infty,$
$\Psi_{3}(r)$ asymptotically approaches the free-wave form, $\Psi
_{3}(r)\approx2\cos(\varphi_{3}-kr).$ Consequently, the main spectral
characteristic of the scattering states, the phase shift, reads $\delta
(k)=(n+1/2)\pi-\varphi_{3},$ where $n$ is an integer. To ensure the correct
energy normalization, Eq. (13) has to be multiplied by the factor $F=\left(
4\pi\sqrt{EC}\right)  ^{-1/2}$ \cite{Landau}. As previously, to ascertain the
phase constant $\varphi_{3},$ one has to suitably adjust Eq. (11) and apply
the boundary condition $\frac{\Psi_{2}^{\prime}(X_{3})}{\Psi_{2}(X_{3})}%
=\frac{\Psi_{3}^{\prime}(X_{3})}{\Psi_{3}(X_{3})}$. Thereafter, one can fix
the normalization factors $N_{2},N_{1}$ and $N_{0}$, using the continuity
conditions for the components of the wave function%
\[
\Psi_{3}(X_{3})=\Psi_{2}(X_{3}),\Psi_{2}(X_{2})=\Psi_{1}(X_{2}),\Psi_{1}%
(X_{1})=\Psi_{0}(X_{1}).
\]

Thus we have explained all details of calculating the stationary wave
functions for an exactly solvable multi-component reference potential. The
described scheme remains the same, independent of how many analytically
different components one introduces. In view of the incomplete knowledge about
the real PEC for Ar$_{2}$, including just four smoothly joined Morse-type
pieces seems quite optimal. In fact, the full discrete energy spectrum
$E_{n}<0$ ($n=0\div6$) can be accurately ascertained with the help of only two
components (OM + RM), because the region $r<X_{2}=3.35$ \AA \ becomes
practically negligible for these calculations. Determining the energy levels
therefore reduces to a very simple zero-finding problem for a function, which
is uniquely determined by the physical boundary conditions $\Psi
_{2}(E,r)\rightarrow0$ as $r\rightarrow0$ and $\Psi_{3}(E,r)\rightarrow0$ as
$r\rightarrow\infty$, complemented with the condition $\frac{\Psi_{2}^{\prime
}(E,r)}{\Psi_{2}(E,r)}=\frac{\Psi_{3}^{\prime}(E,r)}{\Psi_{3}(E,r)}$ for an
arbitrary $r>X_{2}$ (e.g., $r=X_{3}$). These three conditions can be fulfilled
simultaneously only for the true energy eigenvalues, i.e., if $E=E_{n}$.

In Fig. 1 one can see little discrepancies between the original PEC and its
approximant, which could be easily reduced by adding more components. An
important point to discuss in this context is the physically correct
long-range behavior of the potential. Indeed, apart from the artificial
potential barrier, RM approximation seems to be absolutely incompatible with
the attractive inverse power series coordinate dependence, which is expected
at long distances. However, in the framework of the proposed approach, the
"unphysical" nature of the RM approximation and the discrepancies mentioned are
not substantial. First, as can be seen in Table 1, the calculated vibrational
levels for the reference PEC fit with the observed ones \cite{Stoicheff-Ar, CD}
even better than those of the original potential. Thus, adding more components
to the potential is not motivated. Second, in a wide range of actual interest
the RM approximation does not contradict to the inverse power series expansion
(see the inset of Fig. 1). Therefore, since the described approach is reliable
enough, and provides simple analytic solution to the problem, it has been used
in the long distance region as well.

As mentioned, the parameters of the ground state reference potential were kept
unchanged. A large number of scattering ($E>0$) wave functions for this PEC
have been calculated, along with the wave functions of the bound states. These
eigenfunctions are needed as a basis for further calculations, and this basis
should be sufficiently complete to accurately reveal all details of the
Franck-Condon factors for the vibrational levels of the excimer states.

\subsection{Reference potential for the 0$_{u}^{+}$ state of Ar$_{2}^{\ast}$}

Construction of the reference potentials for the excimer states can be
performed in the manner described in the previous subsection. The main
difference is that the parameters of the components are now treated as
variables to be fitted to the experimental data. In addition, the \textit{ab
initio} CCSR(T) daug-cc-pV5Z potential for the 0$_{u}^{+}$\ state
\cite{Fernandez} in the range $r<2.7$ \AA \ has also been used as an input for
the fitting procedure. Therefore, in this region (including the minimum at
2.3893 \AA ) the reference potential practically does not differ from the PEC
calculated by Fern\'{a}ndez and Koch.

As previously, the reference potential was built up of several smoothly joined
Morse-type pieces. Since spectroscopic applications are related to the
excimers' bound states, the short-distance region ($r<2$ \AA ) is of less
interest, and therefore, the reference PEC constructed for the 0$_{u}^{+}$
state does not include any PM components. Thus, the resulting curve shown in
Figs. 2 and 3 has only three constituents with parameters given in Table 2.
The region $r\leq X_{1}$ was approximated by an OM potential, and two RM
components ($r\in\lbrack X_{1},X_{2}]$ and $r\geq X_{2}$, respectively), have
been introduced for the long-distance range where spin-orbit coupling becomes
important. Note that we cannot use the \textit{ab initio }PEC by Fern\'{a}ndez
and Koch for this range, because spin-orbit coupling was ignored in their calculations.

As mentioned, the PM component remained nearly unchanged, while the RM
components have been largely varied, preserving the continuity of the
potential and its derivative, and trying to achieve a good fit with the
experimental data. First, for any intermediate reference PEC, the energy
eigenvalue problem has been solved. As explained in the end of the previous
sub-section, the discrete energy levels can be always found as the zeros of a
characteristic function, which is uniquely determined by the physical boundary
conditions and the continuity requirements (see, e.g., \cite{Selg-2001} and
references therein for more details). Thereafter, the Franck-Condon factors
for all vibrational levels were calculated and integrated over the full energy
range. The results have been compared with the corresponding experimental
data. If needed, the parameters of the reference PEC have been slightly
changed and the whole procedure has been repeated until the desired quality of
the fit was achieved.

Naturally, the correct PEC should reproduce the second continuum and the real
dissociation limit ($E_{a}=11.623592$ eV for the atomic $^{3}$P$_{1}$ level
\cite{NIST}). In addition, the fitting procedure involved the observed level
positions \cite{Stoicheff-Ar} and the intensity patterns from the excimers'
inner turning point region \cite{Moeller}. Unfortunately, only rough
estimations for the important spectroscopic constant $\omega_{e}$ of the
0$_{u}^{+}$ state are available (see, e.g., \cite{Stoicheff-Ar}), but it has
to be close to the values reported for the 1$_{u}$ state ($\omega_{e}=293\pm4$
cm$^{-1}$ \cite{Conrad}, $\omega_{e}=299\pm3$ cm$^{-1}$ \cite{Dube}) and for
the $^{2}\Sigma_{1/2u}^{+}$ state of Ar$_{2}^{+}$ ($\omega_{e}=307\pm0.4$
cm$^{-1}$ \cite{Merkt}). Finally, one has to take account of the known
radiative lifetime for the 0$_{u}^{+}$ constituent of the second continuum,
$\tau=4.20\pm0.13$ ns \cite{Keto}. To achieve agreement with this value, the
transition moment given by Fern\'{a}ndez and Koch has been corrected in the
range $r<3$ \AA \ (see the inset in Fig. 2).

Table 2 and Figs. 3 and 4 demonstrate the obvious success of the described procedure.

\subsection{Reference potential for the 1$_{u}$ state and 1$_{u}$
$\rightarrow$ 0$_{g}^{+}$ transition moment}

The reference potential constructed for the 1$_{u}$ state looks similar to the
0$_{u}^{+}$ PEC just described (see Fig. 2), although the fitting procedure
was slightly different. This time, no \textit{ab initio }PECs can be used for
comparison, but on the other hand, a lot more experimental data are available.
In addition to the level positions and their absolute numbering in the range
$v^{\prime}=23\div31$ \cite{Stoicheff-Ar, Freeman}, the radiative lifetimes
for $v^{\prime}=0$ ($\tau=3.2\pm0.3$ $\mu$s \cite{Keto}) and $v^{\prime
}=24\div30$ \cite{Madej} have been reported, the spectroscopic constant
$\omega_{e}$ is known \cite{Conrad, Dube}, etc.

As for the 0$_{u}^{+}$ state, a reference potential of three components (OM +
RM + RM) has been constructed, but this time all parameters have been varied,
except the equilibrium separation $R_{e}=2.3893$ \AA \ \cite{Fernandez}, which
was kept fixed. The parameters of the PEC and the transition moment can be
fitted independently, since the radiative lifetimes do not depend on the
position of the levels. In its essence, the fitting procedure was the same as
described in the previous subsection. As can be seen from Table 3 and Figs. 3
and 5, a good fit with the experimental data has been achieved for the 1$_{u}$
excimer state as well. The transition moment curve shown in the inset of Fig.
2 is similar to that reported by Madej and Stoicheff \cite{Madej}, but it
falls to zero (1$_{u}$ $\rightarrow$ 0$_{g}^{+}$ transition is forbidden in
the separated-atom limit) more slowly as $r\rightarrow\infty$.

\section{Franck-Condon factors and radiative lifetimes}

Calculation of Franck-Condon factors for the fixed pair of PECs and known
transition moment is a routine but rather time consuming task that has to be
performed very accurately. In the present case, even more computational work
is required because this demanding procedure is used for fitting purposes.
Fortunately, there are some possibilities to reduce the amount of
computations. First, since the ground state reference potential is fixed, one
has to calculate the related wave functions only once. Second, for any
reference PEC under examination, one has to solve the Schr\"{o}dinger equation
only at predefined points, e.g., at the abscissas of the relevant Gaussian
quadrature formula.

Thus, 640 wave functions for the ground state reference PEC have been
calculated in the range $E\in\lbrack0,4.7$ eV] with a variable energy step
from 0.001 to 20 meV. Few examples of these eigenfunctions, all having the
asymptotic form $\Psi(r)\approx2\cos(\varphi_{3}-kr)$ according to Eq. (13),
can be seen in Fig. 6. Of special interest might be the top graph in this
figure, because it illustrates some fundamental findings of the quantum
scattering theory at very low energies, i.e., where $E=Ck^{2}\rightarrow0.$
Since $\tan\varphi_{3}\approx-(kr_{0})^{-1}$ \cite{Davydov} ($r_{0}$ is the
scattering length), and, according to Levinson theorem \cite{Levinson}, the
phase shift $\delta(k)$ $\approx n\pi$ ($n=7$ for the PEC in question) as
$k\rightarrow0,$ one comes to the following expression for the wave function
in this region: $\Psi(r)\approx2k(r-r_{0}).$ This is explicitly demonstrated
by a dotted line in the top graph of Fig. 6, where the vertical dotted line
indicates the position of the characteristic parameter $r_{0}$. Naturally,
such a simple linear coordinate dependence appears only at distances where the
potential well becomes insignificant, i.e., $V(r)\approx0$, but, on the other
hand, $r\ll\frac{\pi}{2k}.$

The scattering wave functions for the ground state reference PEC, along with
the full set of bound state wave functions (only 7 in total), have been used
as the basis to accurately ascertain all details of the Franck-Condon factors.
Throughout the whole domain, the wave functions have been calculated at the
abscissas of the 5-point Gaussian quadrature formula related to the intervals
of 0.02 \AA \ width. Some results of these calculations are shown in Figs. 4,
5, and 7 to 9. In Fig. 7 one can see the Franck-Condon factors for bound-free
transitions from the selected vibrational levels of both 0$_{u}^{+}$ and
1$_{u}$ reference PECs, while the details of the Franck-Condon spectrum from
the highest level ($v^{\prime}=29$) of the 0$_{u}^{+}$ state are demonstrated
in Fig. 8. Note that the actual calculated probability density distributions
are shown on both figures. From these data one can easily ascertain the
probability of spontaneous emission ($p_{v^{\prime}}$) as well as the
radiative lifetime ($\tau_{v^{\prime}}=1/p_{v^{\prime}}$) of the levels. As
explained in handbooks on quantum mechanics (see, e.g., \cite{Davydov}),%
\begin{gather}
p_{v^{\prime}}=\frac{4}{3\hbar^{4}c^{3}}\{\sum_{v^{^{\prime\prime}}%
}(E_{v^{\prime}}-E_{v^{\prime\prime}})^{3}[\int\limits_{0}^{\infty}%
\Psi_{v^{\prime\prime}}(r)\mu(r)\Phi_{v^{\prime}}(r)dr]^{2}\\
+\int\limits_{0}^{\infty}(E_{v^{\prime}}-E)^{3}[\int\limits_{0}^{\infty}%
\Psi(E,r)\mu(r)\Phi_{v^{\prime}}(r)dr]^{2}dE\},\nonumber
\end{gather}
where $E_{v^{\prime}}$ and $\Phi_{v^{\prime}}(r)$ denote the initial (fixed)
energy level and its wave function, while $E_{v^{\prime\prime}}$ and $E$ are
the discrete and continuous energy eigenvalues, respectively, with
corresponding eigenfunctions $\Psi_{v^{\prime\prime}}(r)$ and $\Psi(E,r).$ The
calculations can be conveniently carried out using atomic units, i.e., taking
$\hbar=1,$ the velocity of light $c=137.03604$, and measuring energy in
Hartrees (1 Hartree = 27.2116 eV). Conversion to the SI frequency unit (Hz) is
elementary \cite{Units}: 1 Hz = 24.188843$\cdot10^{-18}$ a.u.

What are actually depicted in Figs. 7 and 8 are the energy integrands of the
second term of Eq. (14) for the selected levels (including the factors
$\frac{4}{3\hbar^{4}c^{3}}$ and $(E_{v^{\prime}}-E)^{3}$). The low-energy part
of the calculated spectra for the 0$_{u}^{+}$ state (see the top graphs in
Figs. 7 and 8) nicely agrees with the experimental results by M\"{o}ller
\textit{et al.} \cite{Moeller}, which confirms their assignment to the inner
turning point region of the excimer's high-lying levels. On the other hand,
this is an evidence of the validity of the \textit{ab initio} PECs by
Fern\'{a}ndez and Koch \cite{Fernandez}. As demonstrated in Fig. 7, the
maximum of the second continuum is expected near 9.8 eV, which is also in full
agreement with experimental observations.

Another interesting result can be seen in the bottom graph of Fig. 8, which is
related to the region where $E\rightarrow0,$ and, consequently, a very small
energy step (0.001 meV) has been used. Namely, since $\Psi(E,r)\rightarrow0$
as $E\rightarrow0$, the high-energy part of the probability density
distribution looks like cut off. Indeed, the squared wave function of the
$v^{\prime}=29$ level, as needed, has 30 maxima, while only 28 maxima are seen
for the related Franck-Condon spectrum in Fig. 8. Naturally, this simply means
that one has to take the bound-bound transitions also into consideration.
These contributions for the selected vibrational levels of both 0$_{u}^{+}$
and 1$_{u}$ excimer states are shown in Fig. 9. The discrete sets of 7 points
($v^{\prime\prime}=0\div6$) there may seem to be located somewhat irregularly,
but in fact, their positions are by no means accidental. On the contrary, as
demonstrated in Figs. 4 and 5, the total integrated transition probability
curves according to Eq. (14) are nice and smooth, although their bound-bound
and bound-free constituents as if show some roughness. Such a behavior of the
total probability is, of course, not only expected but even required. Indeed,
this is nothing else but an indication of the completeness of the basis and
the correctness of normalization of the wave functions. The overall transition
probability for all levels should be exactly the same, if one puts $\mu(r)=1$
and ignores all factors in Eq. (14). Smoothly changing total probabilities
simply reflect the smooth coordinate dependencies of the transition moment.

\section{Conclusion}

For any diatomic system, one can construct an exactly solvable multi-component
reference potential based on the available experimental data. In this paper,
we described a possible strategy to achieve this goal, which is analytically
simple and computationally straightforward. It is based on composing the
reference PECs of several smoothly joined Morse-type potentials, and this was
not an accidental choice. One might think that there are lots of alternatives,
but this is not quite the case. Indeed, for any shape invariant potential one
can easily find the exact solution of the related Schr\"{o}dinger equation
\cite{SUSY}. The point is, however, that they are already the physically
correct linear combinations of special solutions, which can be easily
ascertained only because the specific analytic form of the potential remains
the same in the whole physical domain. The situation changes dramatically, if
the potential consists of several analytically different pieces. Then it is
often possible to easily ascertain one special solution but rarely both of
them. A well-known exception is the piece-wise linear potential possessing two
linearly independent solutions in terms of Airy functions \cite{Wolfram},
nowadays available as standard functions in math-oriented programming
languages. A useful method of solving the Schr\"{o}dinger equation for a
piece-wise linear reference PEC has been worked out long ago by Gordon
\cite{Gordon}.

Another well-known example for which the two linearly independent solutions
can be given in a simple analytic form is the Morse-type PEC analyzed in this
paper. The classical Morse potential \cite{Morse} itself is a reasonable
approximation, but by introducing several components one can get a much more
realistic description. Differently from a piece-wise linear potential, the
components can be smoothly joined, preserving the continuity of the PEC and
its derivative. The two linearly independent solutions of the related
Schr\"{o}dinger equations can be always given in terms of the well-studied
confluent hypergeometric functions, and their correct linear combinations are
determined by the boundary conditions and continuity requirements.

The result of the fitting procedure depends not only on the theoretical
methods and computational techniques applied, but also on the reliability of
the experimental data and the constraints adopted. For example, following
Fern\'{a}ndez and Koch \cite{Fernandez}, we fixed the equilibrium nuclear
separation $R_{e}=2.3893$ \AA \ for both 0$_{u}^{+}$ and 1$_{u}$ excimer
states, but we cannot claim this to be the conclusive value for $R_{e}$
(slightly different results have been reported by other authors). Since the
transitions from the bottom of the excimer states fall into the repulsive wall
of the ground state, the position of the second emission continuum strongly
depends on $R_{e}$. It means that if one assumes a different value for this
parameter, the ground state PEC has to be changed as well. Another issue which
probably needs further confirmation is the absolute numbering of the observed
vibrational levels \cite{Stoicheff-Ar}, because the resulting PECs are rather
sensitive to their changing.

Nevertheless, as we demonstrated, a good fit with the experimental data has
been achieved, and this is a strong argument for the reliability of the PECs
obtained in this paper. The potentials involve a wide range of nuclear
separations ($r\gtrsim1.9$ \AA ) and they can be used to study the most
important spectroscopic features of Ar$_{2}^{\ast}$ excimers, including the
details of their emission continua. In addition, they are expected to be useful
for the analysis of the relaxation dynamics and the time-resolved emission
spectra of the excimers. However, let us recall once again that these
approximate PECs and the relevant transition moments have been deduced on the
relatively modest basis of the available experimental data, following the
relatively "soft" criteria stated in Section II. Hopefully, the results of this
work can stimulate further experimental research to reveal much more details
about the properties of the RG excimers.

\section*{Acknowledgement}

The research described in this paper has been supported by Grant No 5863 from
the Estonian Science Foundation.

\newpage

\section*{Figure captions}

\begin{enumerate}
\item[Fig. 1.] Four-component reference potential (solid line) for the ground
electronic state of Ar$_{2}$ in comparison with two\textit{ ab initio}
potentials. The open circles in both graphs (and in the inset) correspond to
\cite{Fernandez}, while the dotted curve in the upper graph is taken from
\cite{Barker}. The same reference PEC (solid line) is depicted in both graphs, 
but essentially different energy scales are used for them. All components have
the well-known analytic form of the Morse potential, but the ordinary Morse
approximation is used only in the central range $r\in \left[ X_{2},X_{3}%
\right] $ (see the lower graph). Two pseudo-Morse components for the regions 
$r\leq X_{1}$ and $r\in \lbrack X_{1},X_{2}]$, respectively, have been
introduced, while the long-distance range $r\geq X_{3}$ is approximated by a
reversed Morse potential. The inset demonstrates that the RM approximation
(dotted curve) does not contradict to the inverse power series expansion $%
-C_{6}/r^{6}-C_{8}/r^{8}-C_{10}/r^{10}$ (dashed curve). The parameters $%
C_{6},C_{8},$and $C_{10}$ are taken from \cite{Tang}, while the RM
parameters for this specific fit are as follows: $R_{\max }$ = 12.45 \AA , $%
V_{0}$ = -0.1268 meV, $D$ = -1.799E-4 meV, and $\alpha $ = 0.66 \AA $^{-1}$ (%
$V(R)=V_{0}+D\ast \lbrack \exp (-\alpha \ast (R-R_{m}))-1]^{2}$).

\item[Fig. 2.] Reference potentials for the three electronic states of
Ar$_{2}$ examined in this paper. The parameters of the ground-state potential
in the lower graph (the same as in Fig. 1) have been determined from the least
squares fit to the \textit{ab initio} potential by Fern\'{a}ndez and Koch
\cite{Fernandez}, while the curves for the excimer states 0$_{u}^{+}$ and
1$_{u}$ have been constructed with the help of the fitting procedure described
in Sections II B and C, respectively. The corresponding 1$_{u}$ $\rightarrow$
0$_{g}^{+}$ and 0$_{u}^{+}$ $\rightarrow$ 0$_{g}^{+}$ transition moments are
also shown in the upper graph (note that different scales are used for them),
where the dotted line has been taken from \cite{Fernandez}.

\item[Fig. 3.] A more detailed depiction of the three-component reference
potential for the 0$_{u}^{+}$ state (upper graph), and the calculated
vibrational energies for both excimer states in comparison with the observed
data (lower graph). As in the experiments, the position of the levels is
measured relative to the zeroth level of the ground state. The $v^{\prime
}=31$  level of the 1$_{u}$ state was taken from \cite{Freeman}, all other
experimental data are those from \cite{Stoicheff-Ar}.

\item[Fig. 4.] Upper graph: calculated probabilities of bound-free and
bound-bound transitions starting from the vibrational levels of the 0$_{u}%
^{+}$ state. Lower graph: corresponding radiative lifetimes of all levels. The
lifetime of the zeroth level ($\tau_{0}=4.19$ ns) practically coincides with
the experimental result ($\tau_{0}=4.20\pm$ 0.13 ns) from \cite{Keto}.

\item[Fig. 5.] The same as in Fig. 4 but for the 1$_{u}$ excimer state. Again,
the lifetime of the zeroth level ($\tau_{0}=3.15$ $\mu$s) is close to the
experimental one ($\tau_{0}=3.2\pm$ 0.3 $\mu$s) from \cite{Keto}. Good
agreement between the theoretical and experimental \cite{Madej} lifetimes for
the levels $v^{\prime}=24\div30$ is also demonstrated (see Table 3).

\item[Fig. 6.] A selection of the scattering ($E>0$) wave functions for the
ground state (for correct energy normalization they have to be multiplied by
$F=\left(  4\pi\sqrt{EC}\right)  ^{-1/2},$ where $C=\frac{\hbar^{2}}{2m}$).
The upper graph corresponds exactly to the top of the tiny articial potential
barrier (see the explanations in Section II B and Table 1), i.e.,
$E=1.029793\cdot$10$^{-7}$ eV. Due to the extremely small $E=Ck^{2}$, one can
see nearly linear coordinate dependence, $\Psi(r)\approx2k(r-r_{0})$ ($r_{0}$
is the scattering length), in the range where $V(r)\approx0$ but $r\ll
\frac{\pi}{2k}$. The three lower graphs correspond to $E=1$ meV, $E=100$ meV,
and $E=4$ eV, respectively, when the wave function more or less rapidly
achieves its asymptotic free-wave form.

\item[Fig. 7.] Demonstration of the calculated Franck-Condon factors for the
bound-free transitions from the selected vibrational levels of the excimer
states. One can infer that the first and the second emission continua (with
maximum near 9.8 eV) are formed just where expected, while the low-energy part
of the spectrum nicely agrees with the experimental results by M\"{o}ller
\textit{et al.} \cite{Moeller} under selective synchrotron radiation excitation.

\item[Fig. 8.] Demonstration of the details of the Franck-Condon spectrum for
the bound-free transitions from the highest level of the 0$_{u}^{+}$ state
(shown also in Fig. 7). The lower graphs begin exactly at the energies where
the upper graphs end. The bottom graph (where open circles mark the actually
calculated intensities) is cut off at $E=0$, and therefore the overall
spectrum shows only 27 zeros (instead 29). The "missing" part corresponds to
bound-bound transitions.

\item[Fig. 9.] Depiction of the Franck-Condon factors for bound-bound
transitions from the selected vibrational levels of the 0$_{u}^{+}$ (upper
graph) and 1$_{u}$ (lower graph) excimer states of Ar$_{2}$. The related
atomic levels $^{3}$P$_{1}$ and $^{3}$P$_{2}$ are shown by the arrows.
Location of the points corresponding to the highest levels may seem
irregular, but in fact, their positions are determined by rigorous quantum
mechanical sum rules (see also Figs. 4 and 5).
\end{enumerate}

\pagebreak

\begin{table}[ptb]
\caption{Parameters of the reference potential for the ground electronic state
of Ar$_{2}$. The positions of the vibrational levels (in meV, relative to the
bottom of the potential well) are also given.}

\begin{tabular}{p{0.85cm}cccccc}
\hline\hline
$k$ & Type & $V_{k}$ (meV) & $D_{k}$ (meV) & $\alpha _{k}$ (1/\AA ) & $R_{k}$
(\AA ) & Range (\AA ) \\ \hline
0 & PM & -67.83414 & 0.0763369 & 1.708238 & 5.224252 & $r\leq 2.9105$ \\ 
1 & PM & -15.42556 & 0.137657 & 2.293931 & 4.443077 & $r\in \lbrack
2.9105,3.35]$ \\ 
2 & OM & -12.0866 & 11.85914 & 1.729656 & 3.7769 & $r\in \lbrack
3.35,4.4891]$ \\ 
3 & RM & 1.029793E-4 & -1.029793E-4 & 0.6877276 & 12.4891 & $r\geq 4.4891$
\\ \hline
\end{tabular}

\medskip

\begin{tabular}{cccccccc}
\hline
$v^{\prime \prime }$ & 0 & 1 & 2 & 3 & 4 & 5 & 6 \\ \hline
This work & 1.8485 & 5.0741 & 7.6651 & 9.6403 & 10.9256 & 11.7307 & 12.0699
\\ 
Ref. 46 & 1.8127 & 4.9470 & 7.4416 & 9.3224 & 10.6342 & 11.4488 & 11.8691 \\ 
Ref. 30 & 1.8263 & 5.0102 & 7.5619 & 9.4923 & 10.8462 & 11.6918 & - \\ 
\hline\hline
\end{tabular}
\end{table}

\pagebreak

\begin{table}[ptb]
\caption{Parameters of the reference potential for the 0$_{u}^{+}$ state
of Ar$_{2}$. The vibrational energies given in meV are measured relative to the
bottom of the potential well (this work), while the values in cm$^{-1}$ are
transition energies relative to the zeroth level of the ground state, as in
\cite{Stoicheff-Ar}. The calculated spectroscopic constants are as follows:
$\omega _{e}=296.26$ cm$^{-1}$, $D_{e}=6128.3$ cm$^{-1}$.}
\begin{tabular}{p{0.85cm}cccccc}
\hline\hline
$k$ & Type & $V_{k}$ (meV) & $D_{k}$ (meV) & $\alpha _{k}$ (1/\AA ) & $R_{k}$
(\AA ) & Range (\AA ) \\ 
1 & OM & 10864.76 & 1166.681 & 1.66221 & 2.3893 & $r\leq 3.0407$ \\ 
2 & RM & 11624.65 & -0.6649193 & 1.654645 & 4.862433 & $r\in \lbrack
3.0407,4.16]$ \\ 
3 & RM & 11623.59 & -3.021853E-9 & 3.588833 & 7.00 & $r\geq 4.16$ \\ \hline
\end{tabular}

\medskip 
\begin{tabular}{ccccccccc}
\hline
$v^{\prime }$ & 0 & 1 & 2 & 3 & 4 & 5 & 6 & 7 \\ 
$E_{v^{\prime }}$(meV) & 18.2936 & 54.4470 & 90.0222 & 125.0192 & 159.4380 & 
193.2786 & 226.5409 & 259.2250 \\ \hline
$v^{\prime }$ & 8 & 9 & 10 & 11 & 12 & 13 & 14 & 15 \\ 
$E_{v^{\prime }}$(meV) & 291.3308 & 322.8584 & 353.8078 & 384.1790 & 413.9720
& 443.1867 & 471.7515 & 499.7937 \\ \hline
$v^{\prime }$ & 16 & 17 & 18 & 19 & 20 & 21 & 22 &  \\ 
$E_{v^{\prime }}$(meV) & 527.2506 & 553.8589 & 579.4339 & 603.7447 & 626.7231
& 648.2718 & 668.2571 &  \\ 
$E_{v^{\prime }}$(cm$^{-1}$) &  &  &  &  & 92766.9 & 92940.7 & 93101.9 &  \\ 
Exp., Ref. 30 (cm$^{-1}$) &  &  &  &  & 92769.3 & 92935.9 & 93093.5 &  \\ 
\hline
$v^{\prime }$ & 23 & 24 & 25 & 26 & 27 & 28 & 29 &  \\ 
$E_{v^{\prime }}$(meV) & 686.6108 & 703.2166 & 717.9872 & 730.8195 & 741.5951
& 750.1805 & 756.3843 &  \\ 
$E_{v^{\prime }}$(cm$^{-1}$) & 93249.9 & 93383.9 & 93503.0 & 93606.5 & 
93693.4 & 93762.6 & 93812.7 &  \\ 
Exp., Ref. 30 (cm$^{-1}$) & 93241.2 & 93377.6 & 93501.6 & 93610.8 & 93701.3
& - & - &  \\ \hline\hline
\end{tabular}
\end{table}

\pagebreak

\begin{table}[ptb]
\caption{Parameters of the reference potential for the 1$_{u}$ state of
Ar$_{2}$. The vibrational energies given in meV are measured relative to the
bottom of the potential well, while the values given in cm$^{-1}$ are the
transition energies relative to the zeroth level of the ground state.
The corresponding experimental values are from \cite{Stoicheff-Ar}, except
$E_{31}$, which was taken from \cite{Freeman}. The following spectroscopic
constants were obtained for the 1$_{u}$ state:
$\omega _{e}=287.30$ cm$^{-1}$ and $D_{e}=5929.6$ cm$^{-1}$.}
\begin{tabular}{p{0.85cm}cccccc}
\hline\hline
$k$ & Type & $V_{k}$ (meV) & $D_{k}$ (meV) & $\alpha _{k}$ (1/\AA ) & $R_{k}$
(\AA ) & Range (\AA ) \\ 
1 & OM & 10813.17 & 850.0 & 1.888481 & 2.3893 & $r\leq 3.39001$ \\ 
2 & RM & 11548.17 & -0.005519566 & 1.654645 & 6.390011 & $r\in \lbrack
3.39001,5.5]$ \\ 
3 & RM & 11548.35 & -1.515516E-10 & 0.730545 & 20.0 & $r\geq 5.5$ \\ \hline
\end{tabular}

\medskip

\begin{tabular}{cccccccc}
\hline
$v^{\prime }$ & 0 & 1 & 2 & 3 & 4 & 5 & 6 \\ 
$E_{v^{\prime }}$(meV) & 17.7170 & 52.5912 & 86.7190 & 120.1005 & 152.7356 & 
184.6243 & 215.7667 \\ \hline
$v^{\prime }$ & 7 & 8 & 9 & 10 & 11 & 12 & 13 \\ 
$E_{v^{\prime }}$(meV) & 246.1627 & 275.8123 & 304.7156 & 332.8725 & 360.2830
& 386.9472 & 412.8650 \\ \hline
$v^{\prime }$ & 14 & 15 & 16 & 17 & 18 & 19 & 20 \\ 
$E_{v^{\prime }}$(meV) & 438.0364 & 462.4614 & 486.1401 & 509.0725 & 531.2584
& 552.6980 & 573.3912 \\ \hline
$v^{\prime }$ & 21 & 22 & 23 & 24 & 25 & 26 & 27 \\ 
$E_{v^{\prime }}$(meV) & 593.3381 & 612.4904 & 630.8586 & 648.2035 & 664.3098
& 679.0578 & 692.3569 \\ 
$E_{v^{\prime }}$(cm$^{-1}$) &  &  & 92384.2 & 92524.1 & 92654.0 & 92772.9 & 
92880.2 \\ 
Exp., Ref. 30 (cm$^{-1}$) &  &  & 92386.7 & 92524.5 & 92653.1 & 92771.4 & 
92879.1 \\ 
Rad. lifetime (ns) &  &  &  & 182.1 & 169.0 & 159.3 & 152.9 \\ 
Exp., Ref. 31 (ns) &  &  &  & 173 $\pm $ 17 & 162 $\pm $ 14 & 167.2 $\pm $
9.0 & 161.7 $\pm $ 9.4 \\ \hline
$v^{\prime }$ & 28 & 29 & 30 & 31 & 32 & 33 &  \\ 
$E_{v^{\prime }}$(meV) & 704.0696 & 714.0975 & 722.3222 & 728.6210 & 732.8645
& 734.8840 &  \\ 
$E_{v^{\prime }}$(cm$^{-1}$) & 92974.7 & 93055.5 & 93121.9 & 93172.7 & 
93206.9 & 93223.2 &  \\ 
Experiment (cm$^{-1}$) & 92974.8 & 93056.9 & 93123.7 & 93171.0 & - & - &  \\ 
Rad. lifetime (ns) & 149.6 & 149.4 & 152.1 & 158.0 & 167.5 & 185.6 &  \\ 
Exp., Ref. 31 (ns) & 157.4 $\pm $ 9.6 & 155.7 $\pm $ 9.0 & 166 $\pm $ 11 & 
&  &  &  \\ \hline\hline
\end{tabular}
\end{table}

\end{document}